\documentclass[conference]{IEEEtran}
\IEEEoverridecommandlockouts
\usepackage{cite}
\usepackage{amsmath,amssymb,amsfonts}
\usepackage{algorithmic}
\usepackage{graphicx}

\usepackage{listings}
\usepackage{csquotes}

\begin{document}

\title{Bots Don’t Mind Waiting, Do They?\\ Comparing the Interaction With Automatically and Manually Created Pull Requests}

\author{
\IEEEauthorblockN{Marvin Wyrich\IEEEauthorrefmark{1},
Raoul Ghit\IEEEauthorrefmark{2}, Tobias Haller\IEEEauthorrefmark{3} and
Christian Müller\IEEEauthorrefmark{4}}
\IEEEauthorblockA{
Institute of Software Engineering,
University of Stuttgart\\
Stuttgart, Germany\\
\IEEEauthorrefmark{1}marvin.wyrich@iste.uni-stuttgart.de,
\IEEEauthorrefmark{2}raoul.ghit@gmail.com,
\IEEEauthorrefmark{3}tobias\_haller@outlook.de,
\IEEEauthorrefmark{4}muellercn@posteo.de}
}

\maketitle

\begin{abstract}
As a maintainer of an open source software project, you are usually happy about contributions in the form of pull requests that bring the project a step forward.
Past studies have shown that when reviewing a pull request, not only its content is taken into account, but also, for example, the social characteristics of the contributor.
Whether a contribution is accepted and how long this takes therefore depends not only on the content of the contribution.
What we only have indications for so far, however, is that pull requests \textit{from bots} may be prioritized lower, even if the bots are explicitly deployed by the development team and are considered useful.

One goal of the bot research and development community is to design helpful bots to effectively support software development in a variety of ways.
To get closer to this goal, in this GitHub mining study, we examine the measurable differences in how maintainers interact with manually created pull requests from humans compared to those created automatically by bots.

About one third of all pull requests on GitHub currently come from bots.
While pull requests from humans are accepted and merged in 72.53\% of all cases, this applies to only 37.38\% of bot pull requests.
Furthermore, it takes significantly longer for a bot pull request to be interacted with and for it to be merged, even though they contain fewer changes on average than human pull requests.
These results suggest that bots have yet to realize their full potential.

\end{abstract}


%
% -------------------------------------------------------------------------- %
% Introduction
% -------------------------------------------------------------------------- %
%

\section{Introduction}

Software bots are great: they are the interactive and intelligent interface to services designed to improve developers' everyday lives~\cite{Lebeuf:2017:SoftwareBots,Erlenhov:2019:CurrentAndFuture}.
Already in 2017, about 26\% of open source projects have used bots that supported other developers' work, and most developers perceive bots as helpful~\cite{Wessel:2018:PowerOfBots}.
Whether for updating dependencies, automatic bug fixes, or optimizing images, many bots primarily work directly on a project's content and propose their changes to the development team in pull requests for review.

In a recent study, we deployed an autonomous bot to automatically refactor code smells~\cite{Wyrich:2019:Towards} for 41 days in a student development team and qualitatively investigated the perception and acceptance of this bot in an interview study~\cite{Wyrich:2020:Perception}.
Again, our experience was that all participants recognized the usefulness of the bot and expressed a desire to use it in future projects.

This introductory hymn to the success of software bots, however, is followed by another observation from this same study: very few team members felt responsible for the bot, which meant that proposed pull requests by the bot sometimes remained open for one or more weeks~\cite{Wyrich:2020:Perception}.
Human team members had the advantage in finding reviewers for their own pull requests that they could persistently approach other team members and persuade them to review.
The bot did not have this ability to be intrusive, which put it at a disadvantage according to the team members.
As a result, the full potential of the bot was not exploited and fewer code smells were removed from the code base than would have been possible with negligible additional effort.

Since this was a case study, we could not be sure if the finding was solely indicative of a problem with this student team and with this particular bot, or if there was a systemic issue where bots are generally disadvantaged in a project by their limited abilities and by a potential lack of social pressure to engage with the contributing bot.
Most would probably agree that making a human contributor wait feels different than ignoring a bot.

For this reason we aim to complement the findings from the qualitative study with a quantitative investigation.
We conducted a GitHub mining study to exploratively investigate the following \textbf{research question}:
\textit{How does maintainer interaction with pull requests differ between manually and automatically created pull requests?}

%
% -------------------------------------------------------------------------- %
% Related Work
% -------------------------------------------------------------------------- %
%

\section{Related Work}

In this paper, we focus on \textit{DevBots} as defined by Erlenhov et al.~\cite{Erlenhov:2019:CurrentAndFuture}, i.e., those bots that support software development.
Different taxonomies exist to classify bots (e.g.~\cite{Erlenhov:2019:CurrentAndFuture,Lebeuf:2019:Defining}).
Most of the facets listed there do not form a relevant criterion for our work to exclude bots based on them.
For example, which concrete goal the respective bot has, how much it has been given human-like traits or how well it can adapt remains aside at this stage.
Simply put, we are interested in bots that are pull request authors on GitHub. 

The pull-based development model~\cite{Gousios:2014:ExploratoryPull} allows anyone with access to a software project to propose changes to the files in the repository, which are then in most cases asynchronously reviewed by the project's maintainers.
For several years now, bots have also been among the contributors.
In 2017, Wessel et al.~\cite{Wessel:2018:PowerOfBots} analyzed 351 popular GitHub repositories and identified the usage of bots in 93 of them (26\%).
The majority of bots seem to frequently perform similar tasks, mainly updating configuration, documentation and data~\cite{Dey:2020:ExploratoryCommits}.

The main driver for the adoption of bots, according to a study by Erlenhov et al.~\cite{Erlenhov:2020:Practitioners}, is to increase productivity, although this aspect is seen differently by different DevBot users.
We already knew from a study by Meyer et al.~\cite{Meyer:2014:Productivity} that developers perceive their workday as productive if they could complete tasks without significant interruptions.
Interestingly, all DevBot users in Erlenhov's study had to some extent issues with interruptions or noise produced by bots: \enquote{A good bot waits until a developer is ready for feedback}~\cite{Erlenhov:2020:Practitioners}.

Even if bots eventually behave as other project members would like them to, this does not guarantee that their contributions will be treated equally to those of other contributors.
We know that useful contributions are not only evaluated on their content, but also on the social characteristics of the contributor~\cite{Terrell:2017:Gender,Ford:2019:BeyondCode} and that identifying the contributor as bot can be sufficient to observe a negative bias compared to contributions from humans~\cite{Murgia:2016:Among}.

Apart from this, there are also initial findings about which other factors influence the review latency and acceptance of pull requests.
Gousios et al.~\cite{Gousios:2014:ExploratoryPull,Gousios:2015:IntegratorsPersp} found that changes of good quality, to recently modified files and matching the roadmap have a higher chance of being accepted.
Pull requests (PRs) would be closed without merging mainly due to concurrent modifications (27\%), uninteresting changes (16\%) or errors in the implementation (13\%)~\cite{Gousios:2014:ExploratoryPull}.
Tsay et al.~\cite{Tsay:2014:Influence} add that the strength of the social connection between submitter and project manager plays a role in the evaluation of PRs and that well-established projects are more conservative in accepting PRs.
Finally, Wessel et al.~\cite{Wessel:2020:EffectsAdopting} found the adoption of code review bots to increase the number of monthly merged pull requests and to decrease communication among developers.

The time to merge is influenced by many factors, including a developer's previous track record~\cite{Gousios:2014:ExploratoryPull}, size of the project~\cite{Gousios:2014:ExploratoryPull}, the size of the pull request~\cite{Yu:2015:Determinants,Gousios:2015:IntegratorsPersp}, delay to first human response~\cite{Yu:2015:Determinants}, and urgency of the pull request~\cite{Gousios:2015:IntegratorsPersp}.

In our own study~\cite{Wyrich:2020:Perception}, pull requests by a refactoring bot were not processed for long periods of time, according to the team members interviewed, primarily because the bot did not actively seek reviewers sufficiently.
We explained this circumstance, among other things, with \textit{diffusion of responsibility}, a sociopsychological phenomenon in which a person may feel less responsible for actions or inactivity when others are present~\cite{Kassin:2019:SocialPsy}.
There were also no short-term consequences to simply making the bot wait.

The observation that DevBots are not yet realizing their full potential and are occasionally perceived by developers as more distracting than helpful has recently led to a number of publications dealing with successful bot design and bot interaction support, e.g.~\cite{Brown:2019:Sorry,Brown:2020:SorryAgain,Wessel:2020:Inconvenient,Urli:2018:RepairBot}.
Our study is intended to further motivate the desired improvement in bot design and to show, using objectively measurable numbers, how effectively DevBots are currently being used.

%
% -------------------------------------------------------------------------- %
% Methodology
% -------------------------------------------------------------------------- %
%

\section{Methods}

We conducted an exploratory GitHub mining study to compare the interaction with pull requests created by bots and those created by humans using initially defined and objectively measurable comparison criteria.
``Exploratory research cannot provide a conclusive answer to research problems [\ldots], but they can provide significant insights to a given situation''~\cite[p. 64]{Singh:2007:QuantitativeMethods}.
The insights gained in this work should help future work hypothesize how maintainers interact with bots and how bots can be better designed to realize their full potential.

\begin{figure*}[htbp]
    \centering
    \includegraphics[width=1\textwidth]{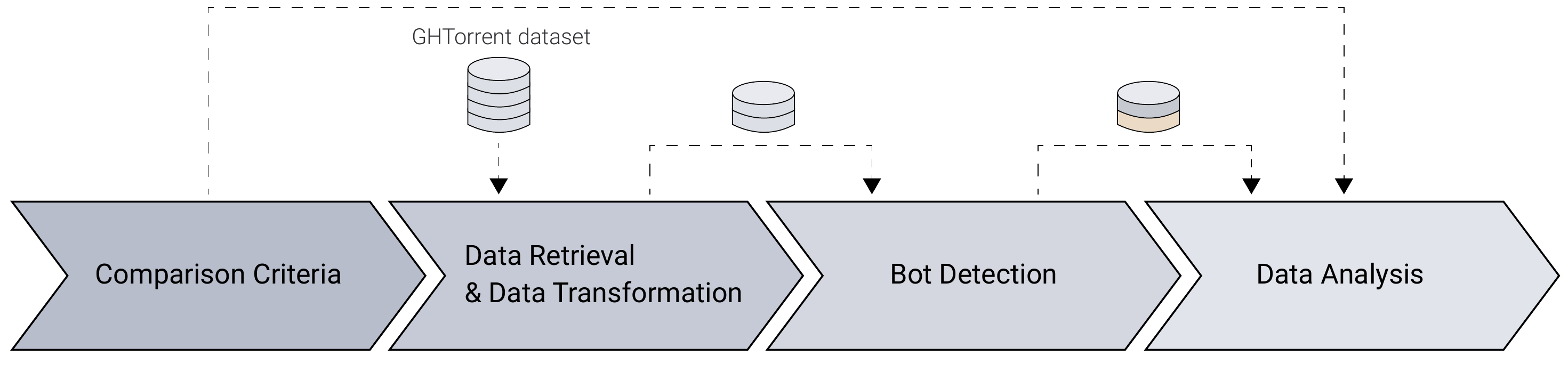}
    \caption{Overview of the research methodology}
    \label{fig:steps}
\end{figure*}

\figurename~\ref{fig:steps} illustrates the steps taken to answer the research question.
Following, each step is explained to justify the design decisions and to be able to replicate the approach.

\subsection{Comparison Criteria}

We chose the comparison criteria based on the assumption motivated in the introduction that pull requests by bots would be accepted less frequently and processed more slowly.
Therefore, the following four proxy variables for maintainer interaction in terms of acceptance and interaction effort with a pull request are considered: acceptance rates, time until first human interaction (i.e., first comment or closing), time until merge by a human and number of comments by humans.

\subsection{Data Retrieval and Data Transformation}

We used GHTorrent~\cite{Gousios:2013:GHtorrent} as data source and limited the dataset to entries from January 1, 2019 to May 31, 2020 to focus on the most recent data and keep the dataset manageable.
From early July 2019 to early January 2020, there is a gap in the GHTorrent data that we cannot explain, but that we also do not suspect will have a large impact on the results.

GHArchive~\cite{GHArchive} was considered as an alternative to GHTorrent.
However, we found that GHArchive had only captured 4,395,331 unique pull requests during the specified time period, compared to 20,623,320 unique pull requests on GHTorrent with the gap mentioned above. We therefore chose the dataset from GHTorrent for this study.

We first filtered the MongoDB dumps for relevant event types and attributes to reduce their size.
The GitHub Events API provides a variety of events, such as pushing to a repository (\texttt{PushEvent}) and creating an issue (\texttt{IssueEvent}).
For our study we only needed all \texttt{PullRequestEvent}, \texttt{IssueCommentEvent}, and \texttt{PullRequestReviewCommentEvent} with their associated attributes.
We then used a self-developed tool\footnote{\label{notePR}Supplemental materials: https://github.com/c-mueller/pr-extractor} to transform these data and load them into a PostgreSQL database.

\subsection{Bot Detection}

Among all GitHub users in our dataset, we now had to identify the bots.
The combination of three approaches turned out to be effective in this context.

First, we looked at the usernames.
There are bots marked as such by the GitHub API with a \texttt{[bot]} suffix that applies only to bots, since square brackets are forbidden characters in usernames.
This results in no false positives, but still a large number of unidentified bots, as this method of bot registration has only recently been implemented~\cite{Integrations,Marketplace} and it may still be more intuitive for developers to simply create a new user account for a bot instead.
With only this approach we identified 1,295 bots that have either created or merged at least one pull request or have at least written one comment on a pull request.

To improve bot detection by username we then considered the expression matching approach proposed by Dey et al.~\cite{Dey:2020:Detecting}. We used the following filter rules: 
1) '\texttt{\%bot}' and '\texttt{\%robot}', 2) '\texttt{\%-bot-\%}' and '\texttt{\%-robot-\%}', 3) '\texttt{bot-\%}' and '\texttt{robot-\%}'.
Exploration has shown that these patterns are very common among bots but quite uncommon among human users.
However, it is clear that this type of bot detection comes with the risk of false positives.
Together with the previously identified bots, we had identified 4,409 bot accounts at this point.

Second, we requested the list of users considered human, who created or merged at least 900 pull requests, or wrote more than 1,500 comments within our specified time frame.
Each profile was manually inspected according to a predefined protocol, and a user was classified as a bot if at least one of the following was true: the profile description states that the user is a bot or the contributions did not look like contributions by a human.
The latter criterion was further refined to arrive at a decision as objectively as possible.
For example, users with more than 10,000 contributions per year or with a continuous contribution pattern were almost always bots, as were users who frequently made identical or very similar contributions.
Through this exploration, we identified an additional 96 bot accounts.

Third, we used the dataset of known bots which resulted from the study by Dey et al.~\cite{Dey:2020:Detecting,Dey:Zenodo}.
Their dataset included all commits from users they identified as bots, with at least 1000 commits, using the commit message, commit association, and author name to identify the bot.
The data was provided in a CSV file, which we processed to obtain the GitHub login names.
Hereby we were able to identify another 149 bot accounts.

During the investigation, we came across seven users who seem to be running a bot using their own personal account.
These hybrid users were added to the bot list, as most of their contributions are most likely created automatically.
The final list of bot accounts contains 4,654 distinct bot login names.

\subsection{Data Analysis}

At that point, we had a relational database that we could run SQL queries on and we knew which accounts in the dataset were bots.
For all four comparison criteria there are two groups of pull requests: those created by bots and those created by humans.
Differences in proportions between these two groups, i.e. acceptance rates, were tested for statistical significance using a Chi-squared test.
To examine differences in the values to the other three comparison criteria, we used a t-test to measure statistical significance and Cohen's d to estimate effect sizes.

\subsection{Data Availability}

We refer the interested reader to our supplemental materials\textsuperscript{1}, which deal in particular with the technical details of data retrieval and transformation and include all SQL queries used.

%
% -------------------------------------------------------------------------- %
% Results
% -------------------------------------------------------------------------- %
%

\bgroup
\def\arraystretch{1.2}
\begin{table}[t]
    \centering
    \caption{Top 12 bot accounts by number of pull requests}
        \begin{tabular}{lr}
            \hline
            \textbf{Bot}                                   & \textbf{\# of PRs} \\
            \hline
            \texttt{dependabot\lbrack bot\rbrack}          & 3,022,938 \\
            %\hline
            \texttt{dependabot-preview\lbrack bot\rbrack}  & 1,222,893 \\
            %\hline
            \texttt{pull\lbrack bot\rbrack}                & 1,005,816 \\
            %\hline
            \texttt{renovate\lbrack bot\rbrack}            & 487,672   \\
            %\hline
            \texttt{pyup-bot}                              & 148,615   \\
            %\hline
            \texttt{greenkeeper\lbrack bot\rbrack}         & 134,218   \\
            %\hline
            \texttt{snyk-bot}                              & 104,628   \\
            %\hline
            \texttt{imgbot\lbrack bot\rbrack}              & 52,255    \\
            %\hline
            \texttt{github-learning-lab\lbrack bot\rbrack} & 39,886    \\
            %\hline
            \texttt{everypoliticianbot}                    & 35,851    \\
            %\hline
            \texttt{depfu\lbrack bot\rbrack}               & 34,408    \\
            %\hline
            \texttt{scala-steward}                         & 32,045    \\
            \hline
        \end{tabular}
    \label{tab:top_bots}
\end{table}
\egroup

\bgroup
\def\arraystretch{1.2}
\begin{table*}[t]
    \centering
    \caption{Pull request (PR) acceptance and interaction statistics by type of author}
        \begin{tabular}{lrrrrrr}
        \hline
        \textbf{PR author} & \textbf{Total} & \textbf{Merged} & \textbf{Avg. Time Until Merge} & \textbf{With Comments} & \textbf{Avg. \# Comments} & \textbf{Avg. Time Until First Interaction}\\
        \hline
        Human & 13,770,280 & 72.53\% & 14 minutes & 23.36\% & 2.52 & 13 minutes \\
        Bot & 6,853,040 & 37.38\% & 10.09 hours & 2.22\% & 0.06 & 12.3 hours \\
        \hline
        \end{tabular}
    \label{tab:results_table}
\end{table*}
\egroup

\section{Results}

To investigate how maintainer interaction with pull requests differs between manual and automatically generated PRs, we analyzed a total of 20,623,320 pull requests created from Jan 1, 2019 to May 31, 2020.
While 1,791,640 human user accounts are responsible for about two-thirds of the pull requests, the remaining third was created by only 4,654 bots.
Even though the average number of created pull requests per bot is very high, more than 80\% of them were created by the four most active bots.
Table~\ref{tab:top_bots} shows the number of pull requests for the top bot contributors in our dataset.
Additionally, pull requests from bots are smaller than those of humans.
Bots add on average about 673 lines and delete 343, while humans add an average of 4,280 lines and delete 1,824.

Table~\ref{tab:results_table} provides an overview of the acceptance and interaction statistics grouped by creator type, i.e. bot and human.
\textit{Total} includes all PRs in the dataset ignoring their current status.
Pull Requests were counted as merged if they were marked as such, regardless of whether human or bot merged it.
The results show that bot pull requests are significantly less likely to be merged (37.38\%) than human created pull requests, which were merged 72.53\% of the time~($p < .001$).

The time it takes humans to merge a pull request also differs significantly between the two groups.
The mean time it takes for a bot pull request to get merged is 10.09 hours, while pull requests created by humans were merged after 14 minutes on average~($p < .001, d = .13$).
Looking at how long it takes for bots to merge pull requests with respect to the type of author, it is the other way around: bot pull requests got merged after only 12 seconds, while human pull requests got merged after an average of 5.42 hours.

When reporting on comments, this includes all human created regular discussion comments under a pull request as well as all human created review comments on a pull request.
Both for human pull requests and bot pull requests, the majority have no comments at all.
Pull requests from bots have significantly, but not substantially, fewer comments compared to human-generated pull requests~($p < .01, d = .0008$).

It takes an average of 13 minutes from pull request creation to the first interaction with a human pull request, whereas it takes significantly longer, 12.3 hours, for a human to respond to a bot pull request~($p < .01, d = .06$).
While the maximum, excluding outliers, for bot pull requests was 298.46 hours to receive a comment, the maximum time for human pull requests was about nine times smaller at about 32.71 hours.

%
% -------------------------------------------------------------------------- %
% Conclusion
% -------------------------------------------------------------------------- %
%

\section{Discussion \& Conclusion}
\label{sec:discussion}

It is interesting to see that one third of all pull requests on GitHub originate from bots.
Unfortunately, only a third of them are accepted and it takes longer than with human pull requests until the first interaction and the merge take place.
While the time variables in their effect size indicate only a small practical difference, the combination of low acceptance rate and significantly longer processing times of bot pull requests suggests that there is potential for improvement, especially since bot pull requests are significantly smaller and could thus be processed faster~\cite{Yu:2015:Determinants,Gousios:2015:IntegratorsPersp}.

To answer the question in the title, no, bots do not mind waiting for their pull request to be merged, but maintainers should care.
Bots can only work effectively if someone feels responsible and interacts with the proposed changes in a timely manner.
The findings of this study provided indications that bots are not yet being used to maximum effect.
Furthermore, evaluation studies of individual bots in terms of their acceptability should take note of our findings on differences in interaction with bots if benchmarking is done with human contributions.
An excellent bot could already be one that is just as well accepted as an average human contributor.

If bots make important contributions, they should receive as much attention as other contributors.
Whether this means developers should change their behavior in dealing with bots, or bots need to be designed differently, and if necessary, be more intrusive to maximize their potential and force the development team to do what is best for them, remains a philosophical question that the bot community can explore in the near future.
Some approaches to help prioritize pull requests, with which some integrators seem to struggle~\cite{Gousios:2015:IntegratorsPersp}, and to accelerate the review process already exist (e.g.~\cite{Maddila:2020:Nudge}).

The main limitation of our study is that we can only describe the differences in interaction, but not explain them.
Additionally, note that in most bot pull requests dependencies were updated while the PRs of the comparison group were more diverse. This difference in goals acts as a confounding factor.
It is possible that bots contribute less urgent changes and are therefore prioritized lower~\cite{Gousios:2015:IntegratorsPersp}.
We also know that for some DevBot users bots are not essential, but only supportive~\cite{Erlenhov:2020:Practitioners}.
As a consequence, bots like dependabot then close their own pull requests after a while because they have become obsolete, leading to a lower acceptance rate.
We leave the assessment of the importance of dependency updates to the developers of the respective projects, but the explanation for the differences in interaction may simply lie in the different content of the pull requests.

Future work can explore the dataset in more depth and, for example, apply the comparison criteria to individual bots to learn if and why individual bots might perform better than others.
Likewise, the bots could be divided into groups according to existing taxonomies~\cite{Erlenhov:2019:CurrentAndFuture,Lebeuf:2019:Defining} and correlations between interaction patterns and the presence of different facets could be examined.
Future research could also investigate whether contributions that are comparable in content are treated the same regardless of the type of contributor, or if the differences seen in our study can be found in this case as well.

%
% -------------------------------------------------------------------------- %
% References
% -------------------------------------------------------------------------- %
%

\bibliographystyle{IEEEtran}
\bibliography{main}

\end{document}